\documentclass[onecolumn,showpacs,preprintnumbers,amsmath,amssymb,showkeys]{revtex4}

\usepackage{graphicx}
\usepackage{dcolumn}
\usepackage{bm}
\usepackage{natbib}

\begin{document}
\draft
\date{\today}
\title{Nonextensive Reaction Rate}
\author{G. B. Ba\u{g}c\i} \thanks{Corresponding Author. Tel.: +1 940 597 8069;fax:+1 940 565 2515.\\}
\email{gbb0002@unt.edu@unt.edu}
\address {Department of Physics, University of North Texas, P.O. Box 311427, Denton, TX 76203-1427,
USA}

\pagenumbering{arabic}

\begin{abstract}
The Kramers' survival probability has been generalized by using
nonextensive formalism. This nonextensive survival probability is
studied in detail and associated Kramers' rate has been calculated
in the high and low viscosity limit. It has been showed that the
proportionality of nonextensive Kramers' rate to the nonextensive
friction term in the high viscosity limit changes to inverse
proportionality in the low viscosity limit. It has also been
observed that friction constant of nonextensive processes is of
rescaled form of the ordinary frictional term. Since the relation
between the ordinary rate and nonextensive rate is found out to be
linear, the Arrhenius nature of the Kramers' rate is preserved. By
using experimental results related to CO rebinding to myoglobin
after photodissociation, we conclude that nonextensivity plays an
important role in protein reactions.
\end{abstract}

  \narrowtext
\newpage \setcounter{page}{1}
\maketitle
\section{\protect\bigskip Introduction}

\noindent \qquad A nonextensive generalization of the standard
Boltzmann-Gibbs (BG) entropy has been proposed by C. Tsallis in 1988
[1-4]. This new definition of entropy is written in terms of a
parameter \textit{q}, a positive constant k which becomes usual
Boltzmann constant in the limit as \textit{q} approaches 1, and
probabilities of microstates. This form of entropy breaks the
additivity of ordinary definition of entropy since for two
independent systems, there is an additional term which consists of
multiplication of these two distinct terms of entropy together with
the parameter (\textit{q}-1). Therefore, the entropic index
\textit{q} is a real number which characterizes the degree of
nonextensivity. When it attains the value 1 as a limiting case,
nonextensive formalism reduces to ordinary Boltzmann formalism which
means that BG statistics is a special case of nonextensive
statistics. Nonextensive formalism is used in systems with
long-ranged interactions, long-ranged memories and systems which
evolves in fractal-like space-time. Even though BG statistics can be
used successfully in investigating extensive systems, physical
systems such as Euler two-dimensional turbulence [5], high energy
collisions [6-9], nematic liquid crystals [10] and \ stellar
polytropes [11] as well as nonisotropic rigid rotator model [12],
Fokker-Planck systems [13,14] can be given as examples for which
nonextensive formalism has been used successfully.

After this brief survey of nonextensive formalism, we turn our
attention to reaction rate problem in the form formulated by Kramers
[15]. He was able to calculate the dependence of escape probability
on viscosity and temperature through his model. Kramers' model has
ben extended by scientists like H\"{a}nggi et al. [16], Montroll and
Schuler [17] for example to consider non-Markovian effects [18]. It
is worth to notice that all these attempts ended up in having a
survival probability which decays exponentially in time. The need of
generalizing Kramers' rate in such a way as to have a nonexponential
decay in time is due to the off-equilibrium condition since this
condition creates genuine power laws as noted by Refs. [19, 20, 21].
Indeed, it has been shown in Ref. [14] that nonlinear Fokker-Planck
equations can result in survival probabilities of the q-exponential
form. Moreover, Plastino et al. [22] considered nonlinear
reaction-diffusion equations with nonlinear diffusion and reaction
term and showed that they possess exact time-dependent particular
solutions of Tsallis' maximum entropy form. Recently, Niven [23]
approached the reaction rate problem in nonextensive formalism from
a different point of view by considering q as the reaction order. In
the next Section, by generalizing survival probability using
nonextensive formalism, we obtain a survival probability which is of
the form of an inverse power law in the asymptotic regime.We also
compare this to some recent experimental findings in protein
rebinding [24, 25, 26]. In Ref. [27], an attempt has been made to
generalize the reaction rate of Kramers through the use of
Mittag-Leffler function where survival probabilities have been in
the form of inverse power law asymptotically. In this study, we
generalize the Kramers' rate using q-exponentials and obtain
survival probabilities of inverse power law. Tsallis et al. in Ref.
[28, 29] uses the same set of data in order to explain fractal
behavior of the experiment through Lyapunov exponents. Our approach
is based on Kramers' model and off-equilibrium conditions which
generate genuine inverse power laws. Within this approach, not only
experimental findings in protein rebinding [24, 25, 26] will be
explained but also the inverse power law behavior found in Ref. [21]
in which a stretched exponential and inverse power law distribution
with different powers are interpolated. The Mittag-Leffler function
requires the same power for stretched exponential and inverse power
law behavior, therefore cannot explain the situation in Ref. [21]
whereas the survival probability of the form of
\textit{q}-exponential can.

\section{Nonextensive Reaction Rates}

Kramers [15] considered a point particle in phase space which is
initially trapped in an asymmetric well under a potential V. In
addition to this, the particle is also assumed to be subject to the
random Brownian forces of the surrounding medium in thermal
equilibrium. The particle can escape over the potential barrier
which means a transition from the well of reactants to the well of
products. Kramers also assumed that the height of the potential
barrier is very large compared with the temperature of the
environment ensuring a slow diffusion process from the well of
reactants to the well of products. It is also assumed that the
potential associated with the well of reactants is given by

\begin{equation}
V(x_{\min })=\frac{1}{2}m(2\pi \omega ^{2})x^{2},
\end{equation}

whereas the potential around the barrier is given by

\begin{equation}
V(x_{\max })=\Delta V-\frac{1}{2}m(2\pi \omega \prime
^{2})(x-x_{\max })^{2},
\end{equation}
The parameters $x_{\min }$ and $x_{\max }$ are the coordinates
whereas $\omega$ and $\omega \prime$ are the corresponding angular
frequencies for the well of reactants and products respectively.
$\Delta V=V(x_{\max })-V(x_{\min })$ is the height of the potential
barrier and m denotes the mass of the particle. The ordinary (i.e.,
extensive) survival probability is then given in terms of usual
exponential function as

\begin{equation}
p(t)=\exp (-rt),
\end{equation}
where r denotes the rate of the process and t is time parameter.

Kramers particularly studied reaction rates for overdamped and
underdamped cases. For the former, he found [15, 30]

\begin{equation}
r\cong\frac{1}{2\pi m}\eta ^{-1}\sqrt{V^{^{\prime \prime }}(x_{\min
})\left\vert V^{^{\prime \prime }}(x_{\max })\right\vert }e^{-\beta
\Delta V},
\end{equation}

and

\begin{equation}
r\cong\eta \beta \Delta Ve^{-\beta \Delta V},
\end{equation}
for the latter, where $\beta $ is (k$_{B}T)^{-1}$ and $\eta $ is the
Brownian friction constant [15]. The double prime indicates the
second derivative with respect to position x. The overdamped case
which is also called the case of large viscosity in Kramers' paper
refers to the case when the effect of the Brownian forces on the
velocity of the particle is much larger than that of the external
force associated with the potential of the well. The underdamped
case makes the assumption that no Brownian forces are present so
that the particle will simply oscillate. This case is also referred
to as the case of low viscosity in Kramers' paper [15]. A brief
explanation for the reaction rate expressions used in this paper for
overdamped and underdamped cases is given in the Appendix. From Eqs.
(4) and (5), it is trivial to observe that r$^{-1}\propto e^{E/T}$
where E$\equiv \Delta V/k_{B}$ i.e., inverse of the Kramers' rate,
both in the high and low viscosity cases, conforms to Arrhenius
activation formula.

Before proceeding with nonextensive formalism, we must inspect Eq.
(3) and observe that ordinary reaction rate can be calculated as
following:

\begin{equation}
r\equiv \frac{p(t=0)}{\widehat{p}(u=0)},
\end{equation}
where $\widehat{p}(u)$ is the Laplace transform of the function
p$(t)$ given by Eq. (3). Laplace transform $\widehat{p}(u)$ of
p$(t)$ is defined by $\widetilde{p}(u)=\int\limits_{0}^{\infty
}p(t)e^{-ut}dt$. Note that when the Laplace variable u is equal to
zero, we have the normalization of the function p$(t)$.

From a mathematical point of view, nonextensive formalism is being
formulated by using the \textit{q}-deformed logarithm and
\textit{q}-deformed exponential which can be given as

\begin{equation}
\ln _{q}x\equiv \frac{x^{1-q}-1}{1-q}\text{ \ \ \ \ \ \ \ \ \ \ \ \
\ \ \ \ \ exp}_{q}x\equiv \lbrack 1+(1-q)x]^{1/(1-q)}.
\end{equation}
As expected, these functions become the usual logarithmic function
and exponential function respectively as \textit{q}$\rightarrow 1.$
Now, we begin by writing ordinary exponential in Eq. (3) as a
q-exponential i.e., we write [14]

\begin{equation}
p_{q}(t)=\exp _{q}(-rt)=[1+(q-1)rt]^{1/(1-q)}.
\end{equation}

p$_{q}(t)$ is normalized as a survival probability in the sense that
it is equal to 1 at t=0, mimicking the exponential case given by Eq.
(3). Then, we rewrite Eq. (6) in the form

\begin{equation}
r_{q}\equiv \frac{p_{q}(t=0)}{\widehat{p}_{q}(u=0)}.
\end{equation}

where the Laplace transform is defined in the same way above. The
Laplace transform $\widehat{p}_{q}$(u) at u = 0 is the integral of
the function p$_{q}(t)$ from zero to infinity i.e., its
normalization. This is obtained as

\begin{equation}
\widehat{p}_{q}(u=0)=\frac{1}{r}\frac{1}{2-q}, \text{q}<2,
\end{equation}
where r is the ordinary rate of the process i.e., $r=r_{q\rightarrow
1}$. The constraint \textit{q} $<2$ has been put since the Laplace
integral in the denominator of Eq. (9) otherwise diverges. Since
p$_{q}(t=0)=1$, we obtain the generalized reaction rate as

\begin{equation}
r_{q}=(2-q)r_{q\rightarrow 1}, \text{q}<2.
\end{equation}

The Eq. (9) is valid for both overdamped and underdamped cases as long as $%
r_{q\rightarrow 1}$ is taken to be of the form in Eqs. (4) and (5)
i.e., the extensive reaction rates for the overdamped and
underdamped cases. For the former case, we get

\begin{equation}
r_{q}=\eta _{q}^{-1}\frac{\sqrt{V^{\prime \prime }(x_{\min })/V^{\prime
\prime }(x_{\max })}}{2\pi m}e^{-\beta \Delta V},
\end{equation}
where
\begin{equation}
\eta _{q}=\frac{\eta }{(2-q)}.
\end{equation}

For the latter, we have

\begin{equation}
r_{q}=\eta _{q}^{\prime }\beta \Delta Ve^{-\beta \Delta V},
\end{equation}

where

\begin{equation}
\eta _{q}^{\prime }=(2-q)\eta .
\end{equation}

From Eqs. (13) and (15), we see that $\eta /\eta
_{\textit{\textit{q}}}=(2-\textit{q})$ and $\eta
_{\textit{q}}^{\prime }/\eta =(2-\textit{q}) $: The nonextensive
friction constants (i.e., $\eta _{\textit{q}}$ and $\eta
_{\textit{q}}^{\prime }$ in the overdamped and underdamped cases
respectively) is rescaled by the same factor (2-\textit{q}), we also
note that nonextensive
formalism gives rise to turnover in the dependence of friction since $%
r_{q}\propto \eta _{q}^{-1}$ and $r_{q}\propto \eta _{q}^{\prime }$
in the overdamped and underdamped cases respectively as can be seen
from Eqs. (12) and (14). This turnover is already inherent in the
Eqs. (4) and (5) which is the extensive theory and we successfully
preserved this form in nonextensive formalism.

The Kramers' theory is also being used for investigating the
chemical reactions in the proteins. However, the related survival
probability in this case is non-exponential. In fact, the experiment
of ligand CO rebinding to myoglobin after photodissociation as
investigated by Iben et al. [24] shows an inverse power law
behaviour in the time asymptotic limit until one reaches a certain
higher critical temperature T$_{c}$. Gl\"{o}ckle and Nonnenmacher
[25] assumed this power to be temperature dependent and equal to
$\alpha (T)=0.41T/120$ to take the change in the protein-solvent
system into account. In Fig. 1, we provide some plots for survival
probability p$_{q}(t)$ for \textit{q} = 2.8, 3.1 and 3.4 which
correspond to these experimental findings [26, 27] for temperature
values T = 160 K, 140 K and 120 K respectively. This plot shows that
as temperature increases, the nonextensivity of the system becomes
less and less dominant.

\begin{figure}
\includegraphics[width=15cm]{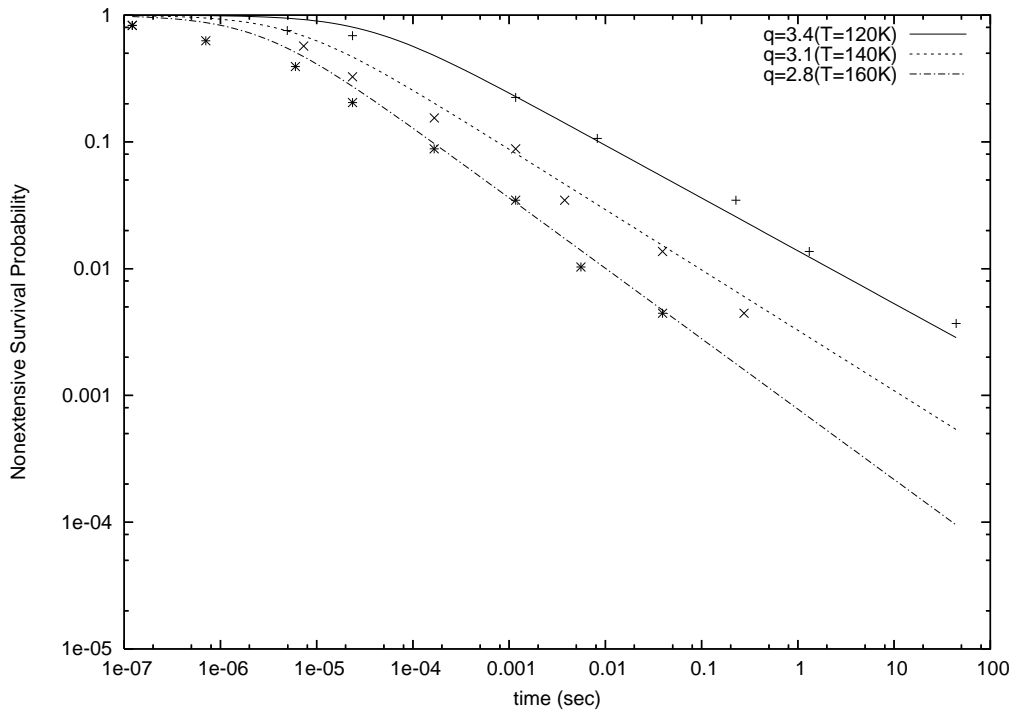}
\caption{\label{fig1} The nonextensive survival probability
p$_{q}(t)$ given by Eq. (8) versus time for the temperatures 120 K,
140 K and 160 K corresponding to the values of \textit{q} equal to
3.4, 3.1 and 2.8 respectively.}
\end{figure}

\section{RESULTS AND DISCUSSIONS}

We have studied nonextensive generalization of Kramers' reaction
rate by writing the survival probability as a q-exponential. We
showed that the dependence of nonextensive Kramers' rate to the
nonextensive friction term in the high viscosity limit changes to
inverse proportionality in the low viscosity limit. In fact, this is
a property of the ordinary Kramers' theory and nonextensive
formalism preserves this important turnover. We calculated
nonextensive reaction rate by making use of Laplace transform and
observed that the relation between the extensive and nonextensive
cases is found out to be linear. Therefore, the Arrhenius nature of
the Kramers' rate is preserved. We then referred to some
experimental data concerning the ligand CO rebinding to myoglobin
after photodissociation. In this experiment, survival probability is
of the form of inverse power law with a power depending on
temperature due to change in the protein-solvent system in the time
asymptotic limit. It has been shown in Fig.1 by plotting
nonextensive survival probability for various temperature values
that some \textit{q} values which are different than 1 correspond to
these distinct cases. We propose this to be a signature of
nonextensivity in the photodissociation of the process of CO
rebinding to myoglobin. Moreover, Fig.1 shows that as temperature
increases, the nonextensivity of the system becomes less and less
dominant. Therefore, this phenemonological picture is in accordance
with the experimental findings which indicate that the survival
probability becomes exponential at a higher temperature. This
transition from power law to exponential can be seen by the
inspection of Fig.1, since as temperature continues to increase, we
expect the nonextensivity parameter \textit{q} to drop to 1 at some
higher critical temperature, which means the transition to
exponential case.In Ref [27], this behavior has been tried to be
explained by the use of Mittag-Leffler function, but we believe that
nonextensive scenario gives a more adequate picture since
Mittag-Leffler forces one to interpolate between the stretched
exponential and inverse power law behavior with same exponent only.
In fact, if one inspects Ref. [21], one immediately sees that it is
a stretched exponential and inverse power law distribution with
different powers to be interpolated. This cannot be done using
Mittag-Leffler function, which requires the same power for stretched
exponential and inverse power law behavior. Our final remark is
about Ref. [28, 29], which uses the same experimental findings as we
did in Fig.1, but the novelty of this paper compared to Ref. [28,
29] lies in the use of different approaches. We tried to generalize
Kramers' rate in a way which will provide survival probabilities of
inverse power law whereas Ref. [28] treats the same subject from the
point of view that the same set of data can be used in order to
explain fractal behavior of the experiment through Lyapunov
exponents.

\section*{ACKNOWLEDGMENTS}

We thank an anonymous referee for very helpful remarks in general
and Ralf Metzler for clarifications regarding the concept of
survival probability through correspondence.

\appendix
\section{}

The reaction rate for the underdamped case is given by Eq. (5) in
our paper and this equation is exactly the same expression in
Kramers' original paper i.e., Eq. (28) in Ref. [15]. The reaction
rate for overdamped case i.e., Eq. (4) in this paper is not found in
Kramers' article [15] but is due to Risken [30]. Kramers' rate for
overdamped case is given by Eq. (17) in Ref. [15] in the following
form

\begin{equation}
r\cong \frac{2\pi \omega \omega ^{\prime }}{\eta }e^{-\beta \Delta
V}.
\end{equation}

In order to see the equality of the equation above and ours given by
Eq. (4), we take derivative of Eqs. (1) and (2) with respect to
position two times and obtain

\begin{equation}
V^{\prime \prime }(x_{\max })=-m(2\pi \omega \prime )^{2}
\end{equation}

and

\begin{equation}
V^{\prime \prime }(x_{\min })=m(2\pi \omega )^{2}.
\end{equation}

Substitution of Eqs. (A2) and (A3) into Eq. (4), we see that Eq.
(A1) follows. Note that we also need to take mass term equal to
unity as Kramers did. Therefore, Eq. (17) in Kramers' paper and our
equation (4) due to Risken is the same.

\end{document}